\documentclass[a4paper,11pt]{article}

\usepackage{amsfonts,latexsym,amsmath,amssymb,amsthm}
\usepackage{graphicx}
\usepackage[francais,english]{babel}
\usepackage[latin1]{inputenc}

\theoremstyle{plain}
\newtheorem{The}{Theorem}[section]

\newtheorem{remark}{\bf Remark}
\newtheorem{Process}{\bf Process}

\newtheorem{Pro}[The]{Proposition}

\title{%An Interacting Particle System Approach for Molecular Dynamics
Equilibrium Sampling From Nonequilibrium Dynamics}
\author{
Mathias Rousset \\
Laboratoire Statistique et Probabilit\'es,
Universit\'e Paul Sabatier\\
118, Route de Narbonne, \\
31062 TOULOUSE Cedex 4 FRANCE\\
{\tt rousset@cict.fr}\\
\\
Gabriel Stoltz \\
CERMICS, Ecole Nationale des Ponts et Chauss\'ees,\\
 6/8 avenue Blaise Pascal,\\
77455 Marne-la-Vall\'ee, France \\
and \\
CEA/DAM Ile-de-France, \\
BP 12, 91680 Bruy\`eres-le-Ch\^atel, France \\
{\tt stoltz@cermics.enpc.fr} \\
\\
}

\begin{document}

\maketitle

\begin{abstract}
We present some applications of an Interacting Particle System (IPS) methodology to the field of Molecular Dynamics. This IPS method allows several simulations of a switched random process to keep closer to equilibrium at each time, thanks to a selection mechanism based on the relative virtual work induced on the system.
It is therefore an efficient improvement of usual non-equilibrium simulations, which can be used to compute canonical averages, free energy differences, and typical transitions paths.
\end{abstract}

%
%  il en faut 4/5 au plus
%
\textbf{Keywords}: %Canonical distribution,
Non-equilibrium molecular dynamics, Interacting Particle System, Genetic Algorithms,
Free energy Estimation. %Feynman-Kac formula.
\vspace{0.5cm}

{\bf AMS:} 65C05, 65C35, 80A10.
\vspace{0.8cm}

Phase-space integrals are widely used in Statistical Physics to relate
the macroscopic properties of a system to the elementary phenomenon
at the microscopic scale ~\cite{FS}. In constant temperature (NVT)
molecular simulations, these integrals often take the form
\begin{equation}
  \label{phase_space_integral}
  \mu(A) = \langle A \rangle = \int_{T^* {\cal M} } A(q,p) \,
d \mu(q,p).
\end{equation}
where ${\cal M}$ denotes the position space (also
called the {\em configuration space}), and $T^*
{\cal M}$ denotes its cotangent space.
A generic element of the position space ${\cal M}$
will be denoted by $q = (q_1, \cdots , q_N)$ and a generic element of the
momentum space by $p = (p_1, \cdots , p_N)$.
We will consider here that ${\cal M} \sim \mathbb{R}^{3N}$ or~$\mathbb{T}^{3N}$ (a torus of dimension $3N$, which arises when using periodic boundary conditions),
and that $T^* {\cal M} \sim \mathbb{R}^{3N} \times \mathbb{R}^{3N}$ or
$\mathbb{T}^{3N} \times \mathbb{R}^{3N}$,
though in general more complicated situations should be considered,
when performing Blue Moon sampling~\cite{CCHK89,CKV05} for example.

The measure $\mu$ is the canonical probability measure
\begin{equation}
  \label{canonical}
  d \mu(q,p) = Z^{-1} \exp (-\beta H(q,p)) \, d q \, d p,
\end{equation}
where $\beta = 1/k_{\rm B}T$ ($T$ denotes the temperature and $k_{\rm{B}}$ the
Boltzmann constant) and where $H$ denotes the Hamiltonian of the
molecular system:
\begin{equation}
  \label{separable_hamiltonian}
  H(q,p) = \frac{1}{2} p^T M^{-1} p + V(q).
\end{equation}
In the above expression, $V$ is the
potential experienced by
the $N$ particles, and $M= \mbox{Diag} (m_1 , \cdots , m_N)$ where
$m_i$ is the mass of the $i$-th particle. The constant $Z$ in~(\ref{canonical}) is the
normalization constant defined as
\[
Z = \int_{ T^{*} {\cal M} } \exp (-\beta H(q,p))
\ d q \, d p.
\]
Some quantities can not be expressed through relations such as~(\ref{phase_space_integral}). One important example is the free energy of a system, defined as
\begin{equation}
\label{free_energy_intro}
F = - \beta^{-1} \ln Z.
\end{equation}

It is though often the case in practice that a straightforward
sampling of $\mu$ is difficult. Indeed, high dimensional systems exhibit
many local minima in which the system remains trapped, especially when
the temperature is low. In those cases, alternative approaches have to
be used, such as those built on the {\it simulated
 annealing}~\cite{KGV83} paradigm. The idea is to switch slowly from an
initial simple sampling problem, to the target sampling problem, through a well chosen interpolation.
This allows to attain deeper local minima, but, due to its nonequilibrium nature, is not efficient as such to sample accurately the target measure.

We mention that variations have been proposed, especially {\it tempering} methods (see~\cite{Iba01} for a review), the most famous being {\it parallel tempering}~\cite{MP92}, which also has an importance sampling version~\cite{Neal01}. These methods consider an additional parameter describing the configuration system (e.g. the temperature), and sample those extended configurations according to some stochastic rules. However, these methods asks for a prior distribution of the additional parameters (for example a temperature ladder in for parallel tempering method), which are usually estimated through some preliminary runs~\cite{Iba01}.

As noted by many authors, simulated annealing strategies can be used to compute exactly ratios of partition function (free energy differences), through an explicit computation of importance weights of nonequilibrium simulations, often referred to as Jarzynski's equality~\cite{jarz, jarzPRL}.
\newline

We present here a complementary approach to the above simulated annealing type
strategies. It is similar to the method of~\cite{HI03}, known as "population Monte-Carlo", in which multiple replicas are used to represent the distribution under study. A weight is associated to each replica, and resamplings are performed at discrete fixed times to avoid degeneracy of the weights. This methodology is widely used in the fields of Quantum Monte
Carlo~\cite{caffarel, moi!} or Bayesian Statistics, where it is referred to as Sequentiel Monte Carlo~\cite{arnaudbook, arnaud}. Note that in the probability and statistics fields,  each simulation is called a 'walker' or 'particle';  we use here the name 'replica', which is more apppropriate to the Molecular Dynamics context.

Our method extends the population Monte-Carlo method to the time-continuous case.
It consists in running $M$~replicas of the system in parallel, resorting typically to a stochastic dynamic, and
considering exchanges between them, according to a certain probability depending on the work done on each system. This procedure can be seen as automatic time continuous resampling, and all replicas have the same weight at any time of the simulation. This method drastically increases the number of significative transitions paths in nonequilibrium simulations. These heuristic explanations are precised in section~\ref{IPS}.  The set of all
replicas (or walkers) is called an 'Interacting Particle System'
(IPS)~\cite{DMseminaire}, and can be seen as a genetic algorithm where the mutation step is the stochastic dynamics considered.
\newline

The article is organized as follows. We first precise classical
simulated annealing type methods in section~\ref{setting}. We then
describe the associated IPS method
in section~\ref{IPS}, as well as its numerical
implementation. Possible applications and some numerical
results are then presented in section~\ref{possible_applications} and~\ref{applications}.

%%%%%%%%%%%%%%%%%%%%%%%%
%
%   Position du pbm
%
%%%%%%%%%%%%%%%%%%%%%

\section{Simulated annealing type methods}

\label{setting}

Consider a family of Hamiltonian functions
$H_{\lambda}: T^*{\cal M} \rightarrow \mathbb{R}$ indexed by a parameter
$\lambda\in [0,1]$.
The corresponding Hamiltonian dynamics are
\begin{equation}
  \label{eqm_ham}
  \left \{
  \begin{array}{c@{ = }c}
    \displaystyle \frac{dq}{dt} & \displaystyle \frac{\partial H_\lambda}{\partial p}, \\
    \displaystyle \frac{dp}{dt} & \displaystyle - \frac{\partial H_\lambda}{\partial q}.\\
  \end{array}
  \right.
\end{equation}
The family $(H_\lambda)_{\lambda \in [0,1]}$ indexes a path between the original state described by a Hamiltonian $H_0$ and the final state charaterized by a Hamiltonian $H_1$.
A canonical probability measure $\mu_\lambda$ can be associated to each Hamiltonian~$H_\lambda$~:
\begin{equation}
\label{canonical_lambda}
d\mu_{\lambda}(q,p) = \frac{1}{Z_\lambda} {\rm e}^{-\beta H_{\lambda}(q,p)} \, dq \, dp,
\end{equation}
where the normalizing constant $Z_\lambda$ is 
\[
Z_\lambda = \int_{\cal M} {\rm e}^{-\beta H_{\lambda}(q,p)} \, dq \, dp.
\]
We will also denote by $F(\lambda) = - \beta^{-1} \ln Z_\lambda$ the corresponding free energy.

We wish to sample according to $d\mu_{1}$, from an initial, easily obtained sample of
$d\mu_{0}$. For each replica of the previous sample, the corresponding
configuration of the system is brought slowly to the end state along a path $(\lambda(t))_{t \in [0,T]}$ for a time $T > 0$. The final sample of configurations is hopefully close to $d\mu_{1}$.

Typically, we can consider $H_\lambda(q,p) = (1-\lambda) H_0(q,p) + \lambda H_1(q,p)$ (e.g. when performing a change temperature $H_0(q,p) = H(q,p), \ \ H_1(q,p) = \frac{\beta'}{\beta} \, H(q,p)$).
It can also represent a modification of the potential, sometimes called 'alchemical transition' in the physics and chemistry litterature. The folding of a protein could be studied this way for example, by setting initially all the long-range interactions to zero, whereas the final state corresponds to a Hamiltonian were all interactions are set on.
In this case,
\[
H_\lambda(q,p) = \frac{1}{2} \, p^T M^{-1} p + V_\lambda(q).
\]

%-------------------
% Gabriel 22/10/05
The simulated annealing like strategies can also be extended to the reaction coordinate case~\cite{LRS06}. In this case, the initial and the final state are indexed by some order parameter $z(q)$. 

\subsection{Markovian nonequilibrium simulations}
The usual way to achieve this method is to perform a time inhomogeneous
irreducible Markovian dynamic
\begin{equation}
\label{e:fdyn}
t \mapsto X^{\lambda(t)}_{t}, \quad X_0^{\lambda(0)} \sim \mu_{0},
\end{equation}
for $t\in [0,T]$, and a smooth schedule $t \mapsto \lambda(t)$ verifying
$\lambda(1)=0$ and $\lambda(T)=1$, and such that
 for all given $\lambda \in [0,1]$, the Boltzmann distribution $d\mu_{\lambda}$ is invariant
  under the dynamics $t \mapsto X_{t}^{\lambda}$.\\

The variable $x$ can represent the whole degrees of freedom $(q,p)$ of
the system, or only the configuration part $q$. Depending on the context, the invariant measure $\mu$ will therefore be the canonical measure~(\ref{canonical}),
or its marginal with respect to the momenta, which reads
\begin{equation}
\label{canonical_overdamped_lambda}
d\tilde{\mu}_{\lambda}(q) = \frac{1}{\tilde{Z}_\lambda}
{\rm e}^{-\beta V_{\lambda}(q)} \, dq,
\end{equation}
with
\[
\tilde{Z}_\lambda = \int_{\cal M} {\rm e}^{-\beta V_{\lambda}(q)} \, dq.
\]
When we do not wish to precise further the dynamics, we simply call
$d\mu_\lambda(x)$ the invariant measure, and $x$ the configuration of the system.
The actual invariant measure should be clear from the context.

For all $t\in[0,T]$, the dynamic~\eqref{e:fdyn} will be usefully characterized by its infinitesimal
generator $L_{\lambda(t)}$, defined on a domain of continuous bounded test functions $\varphi$ by:
$$ L_{\lambda(t)}(\varphi)(x) = \lim_{h\rightarrow 0^{+}}\frac{1}{h}\big(\mathbb{E}(\varphi(X^{\lambda(t+h)}_{t+h})|X_{t}^{\lambda(t)}=x)-\varphi(x)\big)$$
The invariance of $\mu_{\lambda(t)}$ under the instantaneous dynamics can be expressed through the balance condition:
\begin{equation}
  \label{e:invariance}
  \forall \varphi,  \quad \mu_{\lambda(t)}(L_{\lambda(t)}(\varphi)) = 0.
\end{equation}
The dynamics we have in mind are (for a {\emph fixed} $\lambda \in [0,1]$):
\begin{itemize}
\item The hypo-elliptic Langevin dynamics on $T^*{\cal M}$
\begin{equation}
\label{e:Pqdyn}
\left \{ \begin{array}{cl}
d q^{\lambda}_{t} & = \displaystyle \frac{\partial H_\lambda}{\partial p}(q^\lambda_t,p^\lambda_t) \, dt, \\
d p^{\lambda}_{t} & = \displaystyle -\frac{\partial H_\lambda}{\partial q}(q^\lambda_t,p^\lambda_t) \, dt - \xi
M^{-1} p_{t}^{\lambda} \, dt + \sigma \, d W_t, \\
\end{array} \right.
\end{equation}
where $W_t$ denotes a standard $3N$-dimensional Brownian motion.
The paradigm of Langevin dynamics is to introduce in the Newton
equations of motion~(\ref{eqm_ham}) some fictitious brownian forces modelling fluctuations,
balanced by viscous damping forces modelling dissipation.
The parameters $\sigma, \xi > 0$ represent the magnitude of the
fluctuations and of the dissipation respectively, and are linked by
the fluctuation-dissipation relation:
\begin{equation}
\label{FDR}
\sigma = (2 \xi/\beta)^{1/2}.
\end{equation}
Therefore, there remains one adjustable parameter in the model.
The infinitesimal generator is given by:
\[
L_{\lambda}\varphi = \frac{\partial H_\lambda}{\partial
  p}\cdot \nabla_{q}\varphi - \frac{\partial H_\lambda}{\partial
  q}\cdot \nabla_{p}\varphi - \xi
M^{-1} p \cdot \nabla_{p}\varphi  + \frac{\xi}{\beta} \Delta_{p}\varphi.
\]

\item The elliptic overdamped Langevin dynamic\footnote{
This dynamics is actually known as the 'Langevin dynamic' in the probability and statistics fields. We adopt here the physical names of these stochastic processes, which are more natural when dealing with molecular dynamics.
} in the configuration space~${\cal M}$:
\begin{equation}
\label{e:Pdyn}
d q^{\lambda}_t = -\nabla V_{\lambda}(q^{\lambda}_{t}) \, dt
+ \sigma \,
\end{equation}

where the magnitude of the random forcing is given here by
\[
\sigma = \sqrt{\frac{2}{\beta}}.
\]
The corresponding infinitesimal generator is given by:
\[
L_{\lambda}\varphi = \frac{1}{\beta} \Delta_{q}\varphi - \nabla
V_{\lambda}(q) \cdot \nabla_{q}\varphi.
\]
Let us remark that the overdamped Langevin dynamic~(\ref{e:Pdyn}) is
obtained from the Langevin dynamic~(\ref{e:Pqdyn}) by
letting the mass matrix $M$ go to zero and by setting $\xi=1$, which
amounts here to rescaling the time.
\end{itemize}
It is well known that, for a {\emph fixed} $\lambda \in [0,1]$, these dynamics are ergodic under mild assumptions on the potential $V$~\cite{gaby}.

When the schedule is sufficiently
slow, the dynamics is said quasi-static, and the law of the process
$X^{\lambda(t)}_{t}$ is assumed to
stay close to its local steady state throughout the transformation.
As said before, this is out
of reach at low temperature (more precisely, large deviation results~\cite{fried} ensure that
the typical escape time from metastable states grows exponentially fast with $\beta$,
which implies quasi-static transformations to being exponentially slow
with $\beta$).

It is therefore interesting to consider approaches built on the simulated annealing formalism, but able to deal with reasonably fast transition schemes.

\subsection{Importance weights of nonequilibrium simulations.}
\label{FK}

%------------------------
% MODIFS 19/10/05 GABRIEL
For a given nonequilibrium run $X^{\lambda(t)}_{t}$ we denote by
\begin{equation}
\label{work}
W_{t} = \int_{0}^{t}\frac{\partial H_{\lambda(s)}}{\partial
  \lambda}(X^{\lambda(s)}_{s}) \lambda'(s) \, ds
  \end{equation}
the out of equilibrium
virtual work induced on the system on the time schedule $[0,t]$. The quantity
$W_{t}$ gives the importance weights of nonequilibrium simulations with respect to the target equilibrium distribution. Indeed, it was shown in~\cite{jarz} that 
\begin{equation}
\label{e:jar-rel}
\mathbb{E}({\rm e}^{-\beta W_{t}}) = {\rm e}^{-\beta(F(\lambda(t))-F(0))}.
\end{equation}
This equality is known as the Jarzynski's equality, and can be derived through a 
Feynman-Kac formula~\cite{HS01}.
Differentiating the un-normalized Boltzmann path $t \mapsto \Pi_{\lambda(t)}(dx)= {\rm e}^{-\beta H_{\lambda(t)}(x)} \, dx$ with respect to $t$:
\[
\partial_{t} \Pi_{\lambda(t)}(\varphi) =
-\Pi_{\lambda(t)} \left ( \beta \frac{\partial H_{\lambda(t)}}{\partial
\lambda} \lambda'(t) \varphi \right ),
\]
and using the balance condition~(\ref{e:invariance}),
\begin{equation}
\label{e:diffPi}
\partial_{t} \Pi_{\lambda(t)}(\varphi) =
\Pi_{\lambda(t)} \left ( L_{\lambda(t)}(\varphi) - \beta\frac{\partial
H_{\lambda(t)}}{\partial \lambda} \lambda'(t) \varphi \right ),
\end{equation}
it follows
\begin{equation}
\label{e:FK}
\mu_{\lambda(t)}(\varphi) \propto \frac{\Pi_{\lambda(t)}(\varphi)}{\Pi_{\lambda(0)}(1)}
= \mathbb{E} \left ( \varphi(X^{\lambda(t)}_{t}) {\rm e}^{-\beta W_{t}} \right ).
\end{equation}
Therefore, taking $\varphi=1$, 
\begin{equation}
\mathbb{E}({\rm e}^{-\beta W_{t}}) = {\rm e}^{-\beta(F(\lambda(t))-F(0))}.
\end{equation}
Jensen's inequality then gives
\begin{equation}
\mathbb{E}(W_{t}) \geq F(\lambda(t))-F(0).
\end{equation}
This inequality is an equality if and only if the transformation is
quasi-static on $[0,t]$; in this case the random variable $W_t$ is actually constant and equal to $\Delta F$.  When the evolution is reversible, this means that equilibrium
is maintained at all times.

%------------------------
% MODIFS 19/10/05 GABRIEL
%
%  Question : je ne comprends pas d'où vient l'intégration thermo que tu écris, en tout cas
%  comment elle provient de la Feynman-Kac...??
%
%  Réponse: c'est juste pour introduire le concept de force et la notation associée, je l'ai déplacé au paragraphe d'après.
%

As an improvement, we will use the importance weights $W_t$ to perform a selection between replicas.
%----------------------------

%%%%%%%%%%%%%%%%%%%%%%%%%%%%%%%%
%
%  IPS
%
%%%%%%%%%%%%%%%%%%%%%%%%%%%%%%%%

\section{The Interacting Particle System method}
\label{IPS}

Our strategy is inspired by the re-sampling methods in Sequential Monte Carlo
(SMC) literature~\cite{arnaud, moi/arnaud}.
In this time continuous context, the idea is to replace importance weights of simulations performed in parallel, by a selection operation between replicas.

%------------------------
% MODIFS 19/10/05 GABRIEL
We first describe the IPS approximation in section~\ref{discretization}, as well as convergence
results of the discretized measure to the target measure. A justification through a mean-field interpretation is then presented in section~\ref{IPSmethod}. The numerical implementation of the IPS method is eventually discussed in section~\ref{numerical_implementation}.
%------------------------

\subsection{The IPS ans its statistical properties}
\label{discretization}
Recall that the potential of mean force defined by
$$ \mathcal{F}_{\lambda(t)}=\mu_{\lambda(t)} \left ( \frac{\partial
H_{\lambda(t)}}{\partial \lambda} \right )$$
is the average force applied to the system during an infinitely slow transformation.
It can be used in a thermodynamic integration to compute free energy differences:
\begin{equation}\label{e:it}
F(1)-F(0)=\int_{0}^{T} \mathcal{F}_{\lambda(t)}\lambda'(t) \, dt.
\end{equation}
The first step is to rewrite the Feynman-Kac formula~\eqref{e:FK} by introducing a dichotomy when a replica is receiving either excess or deficit work compared to the potential of mean force.

To this end, we define respectively the excess and deficit force, and the excess and deficit work as
\begin{equation}
f^{\rm ex}_{t}(x) =
\left (\frac{\partial H_{\lambda(t)}}{\partial \lambda} -\mathcal{F}_{\lambda(t)}\right ) ^{+}\!\!(x), \quad  f^{\rm de}_{t}(x) =
\left (\frac{\partial
H_{\lambda(t)}}{\partial\lambda} -\mathcal{F}_{\lambda(t)}\right ) ^{-}\!\!(x)  \nonumber
\end{equation}
\begin{equation}\label{e:ex/de_work}
W^{\rm ex}_{t} = \int_{0}^{t} f^{\rm ex}_{s}(X^{\lambda(s)}_{s}) \lambda'(s) \ ds, \quad W^{\rm de}_{t} =  \int_{0}^{t} f^{\rm de}_{s}(X^{\lambda(s)}_{s}) \lambda'(s) \ ds,
\end{equation}
and rewrite
\begin{equation}
\label{e:FKdicho}
\mu_{\lambda(t)}(f)=\frac{\mathbb{E} \left ( f(X^{\lambda(t)}_{t}) {\rm e}^{-\beta (W^{\rm ex}_{t}-W^{\rm de}_{t})} \right )}{\mathbb{E} \left({\rm e}^{-\beta (W^{\rm ex}_{t}-W^{\rm de}_{t})} \right)}.
\end{equation}

We now present the particle interpretation of~(\ref{e:FKdicho}) enabling a numerical computation through the use of empirical distributions. Consider $M$ Markovian systems described by variables $X_t^k$ ($0 \leq k \leq M$). We approximate the virtual force by
\[
\mathcal{F}^{M}_{\lambda(t)} = \frac{1}{M} \sum_{k = 1}^{M}
\frac{\partial H_{\lambda(t)}}{\partial\lambda}(X^{k}_{t}),
\]
and the Boltzmann distribution by
\[
d\mu^{M}_{\lambda(t)}(x) = \frac{1}{M} \sum_{k=1}^{M} \delta_{X^{k}_{t}}(dx),
\]
which are their empirical versions. This naturally gives from definitions~\eqref{e:ex/de_work} empirical approximations of excess/deficit forces $f^{M,{\rm ex/de}}_{t}$ and works $W^{k, {\rm ex/de}}_{t}$.

The replicas evolve according to a branching process whith the following stochastic rules
(see~\cite{moi!,moithese} for further details):

\begin{Process}
\label{p1}
Consider an initial distribution $(X^1_0,\dots,X^M_0)$ generated from
$d\mu_0(x)$. Generate idependent times $\tau^{k,b}_{1},\tau^{k,d}_{1}$ from an exponential law of mean $\beta^{-1}$ (the upperscripts $b$ and $d$ refer to 'birth' and 'death' respectively), and initialize the jump times $T^{b/d}$ as $T^{k,d}_{0} = 0, T^{k,b}_{0} = 0$.

For $0 \leq t \leq T$,
\begin{itemize}
\item Between each jump time, evolve independently the replicas $X_t^k$
  according to the dynamics~\eqref{e:fdyn};
\item At random times $T^{k,d}_{n+1}$ defined by
\[ W^{k,{\rm ex}}_{T^{k,d}_{n+1}}-W^{k,{\rm ex}}_{T^{k,d}_{n}}= \tau^{k,d}_{n+1},
\]
an index $l \in \{1,\dots,M\}$ is picked at random,
and the configuration of the $k$-th replica is replaced by the configuration of the $l$-th replica. A time $\tau^{k,d}_{n+2}$ is generated from an exponential law of mean $\beta^{-1}$;
\item At random times $T^{k,b}_{n+1}$ defined by
\[ W^{k,{\rm de}}_{T^{k,d}_{n+1}}-W^{k,{\rm de}}_{T^{k,d}_{n}}
 = \tau^{k,b}_{n+1},
\]
an index $l \in \{1,\dots,M\}$ is picked at random,
and the configuration of the $l$-th replica is replaced by the configuration of the $k$-th replica. A time $\tau^{k,b}_{n+2}$ is generated from an exponential law of mean $\beta^{-1}$.
\end{itemize}
\end{Process}

The selection mechanism therefore favors replicas which are sampling
values of the virtual work $W_{t}$ lower than the empirical average.
The system of replicas is 'self-organizing' to keep closer to a quasi-static
transformation.

In \cite{DMseminaire,moi!}, several convergence results and
statistical properties of the replicas distribution are proven.
They are summarized in the following proposition:
\begin{Pro}
\label{p:ips}
Assume that $(t,x) \mapsto \frac{\partial H_{\lambda(t)}}{\partial\lambda}(x)$ is a continuous bounded function on $[0,T] \times  T^*{\cal M}$
(or $[0,T] \times {\cal M}$ in the case of overdamped Langevin dynamics), and that the dynamic~\eqref{e:fdyn} is Fellerian and irreducible. Then for all test function $\varphi$ and any $t \in [0,T]$,
\begin{itemize}
\item The estimator 
\begin{equation}\label{e:unbias}
\mu_{\lambda(t)}^{M}(\varphi) \exp \left ( -\beta\int_{0}^{t}
\mathcal{F}^{M}_{\lambda(s)} \lambda'(s) \, ds \right )
\end{equation}
is an unbiased estimator of~\eqref{e:FK};
\item the estimator $\mu_{\lambda(t)}^{M}(f)$ is an asymptotically normal estimator of
$\mu_{\lambda(t)}(f)$, with bias and variance of order $M^{-1}$.
\end{itemize}
\end{Pro}

The proof follows from Lemma~3.20, Proposition~3.25 and Theorem~3.28
of~\cite{DMseminaire} (see also~\cite{moi!,moithese} for further details).

The unbiased estimation of un-normalised quantities is a very usual property in importance sampling type strategies.
In discrete time SMC methods, it comes from the fact that at each time step, when operating re-sampling, each replica branches
with a number of offsprings proportional in average to its importance weight. Unfortunately, there is no simple justification of this fact for time continuous IPS since the branching phenomenon is diluted in the continuity of time.

\subsection{Justification through a mean-field limit}
\label{IPSmethod}

In order to prove the consistency of the IPS approximation, we consider the ideal setting where the number of replicas goes to infinity ($M\rightarrow +\infty$). This point of view is equivalent to a mean-field or Mc Kean interpretation of the IPS (denoted by the superscript 'mf'). In this limit, the behavior of any single replica, denoted by $X_{t}^{\rm mf}$, is then independent from any finite number of other ones. We shall consider the mean field distribution and force:
\begin{eqnarray*}
\textrm{Law}(X_{t}^{\rm mf})&=&\mu_{t}^{\rm mf}, \\
\mathcal{F}^{\rm mf}_{t}&=&\mu_{t}^{\rm mf} \left ( \frac{\partial H_{\lambda(t)}}{\partial\lambda} \right ). \\
\end{eqnarray*}
The mean field excess/deficit force $f_{t}^{\rm mf,ex/de}$ and works $W_{t}^{\rm mf,ex/de}$ are defined as in~\eqref{e:ex/de_work}.

In view of Process~\ref{p1}, the stochastic process $X^{\rm mf}_{t}$ is a jump-diffusion process which evolves according to the following stochastic rules (some facts about pure Markov jump processes are recalled in the Appendix):
\begin{Process}
\label{p2}
Generate $X^{\rm mf}_{0}$ from $d\mu_0(x)$. Generate idependent clocks
$(\tau^{b}_{n},\tau^{d}_{n})_{n \geq 1}$ from an exponential law of mean $\beta^{-1}$ (the upperscripts $b$ and $d$ refer to 'birth' and 'death' respectively), and initialize the jump times $T^{b/d}$ as $T^{d}_{0} = 0, T^{b}_{0} = 0$.

For $0 \leq t \leq T$,
\begin{itemize}
\item Between each jump time, $t \mapsto X^{\rm mf}_{t}$ evolves
  according to the dynamic~\eqref{e:fdyn};
\item At random times $T^{d}_{n+1}$ defined by
\[
W^{\rm mf,ex}_{T^{d}_{n+1}}-W^{\rm mf,ex}_{T^{d}_{n}} = \tau^{d}_{n+1},
\]
the process jumps to a configuration $x$, chosen according to the probability measure
$d\mu^{\rm mf}_{T^{d}_{n+1}}(x)$;
\item At random times $T^{b}_{n+1}$ defined by
\[
\mathbb{E}(W^{\rm mf,de}_{t})|_{t=T^{b}_{n+1}}-\mathbb{E}(W^{\rm mf,de}_{t})|_{t=T^{b}_{n}} = \tau^{b}_{n+1},
\]
the process jumps to a configuration $x$,
chosen according to the probability measure proportional to $\displaystyle f^{\rm mf,de}_{T^{b}_{n+1}}(x) d\mu_{\lambda(T^{b}_{n+1})}(x)$.
\end{itemize}
\end{Process}

\begin{remark}
Note that, in the treatment of the deficit work, we take in Process~\ref{p2} the point of view of the jumping
%------------------------
% MODIFS 19/10/05 GABRIEL -> c'est un peu plus neutre...!
%killed
replica; whereas in Process~\ref{p1}, we take the point of view of the attracting replica which induces a branching.
\end{remark}

From the above probabilistic description, we can derive the Markov generator of the mean-field process, given by the sum of a diffusion and a jump generator:
$$ L^{\rm mf}_{t} = L_{\lambda(t)} + J_{t,\mu^{\rm mf}_{t}}, $$
where the jump generator $J_{t,\mu^{\rm mf}_{t}}$ is defined as
\[
J_{t,\mu^{\rm mf}_{t}}(\varphi)(x) = \beta \lambda'(t)
\int_{\cal M} (\varphi(y)-\varphi(x)) (f^{\rm mf,ex}_{t}(x) + f^{\rm mf,de}_{t}(y)) \mu^{\rm mf}_{t}(dy).
\]
The non-linear evolution equation of the mean-field distribution $\mu^{\rm mf}_{t}$ is then
\begin{equation*}
\label{e:nonlinear}
\partial_{t} \mu^{\rm mf}_{t}(\varphi) = \mu^{\rm mf}_{t} \left ( L_{\lambda(t)}(\varphi) + J_{t,\mu^{\rm mf}_{t}}(\varphi) \right ),
\end{equation*}
which gives, by a straightforward integration,
\begin{equation*}
\label{Boltz_diff_non_norm}
\partial_{t} \mu^{\rm mf}_{t}(f) = \mu^{\rm mf}_{t} \left ( L_{\lambda(t)}(\varphi) + \beta \left ( \mathcal{F}^{\rm mf}_{t} - \frac{\partial H_{\lambda(t)}}{\partial \lambda} \right ) \lambda'(t) f \right ),
\end{equation*}
so that finally
\begin{equation*}
\partial_{t}( \mu^{\rm mf}_{t}(\varphi) {\rm e}^{-\beta \int_{0}^{t}\mathcal{F}^{\rm mf}_{s} \ ds } ) = \mu^{\rm mf}_{t} \left( L_{\lambda(t)}(\varphi) - \beta \frac{\partial H_{\lambda(t)}}{\partial \lambda}\lambda'(t) \varphi \right).
\end{equation*}
Since the {\it un-normalized} Boltzmann distribution is the unique solution of the above linear equation (see \ref{e:diffPi}), we obtain the following identities:
\begin{equation}
\mu^{\rm mf}_{t} = \mu_{\lambda(t)}, \quad
\mathcal{F}^{\rm mf}_{t} = \mathcal{F}_{\lambda(t)}, \quad
f_{t}^{\rm mf,ex/de} = f_{\lambda(t)}^{\rm ex/de}.
\end{equation}
This proves the consistency of the IPS approximation scheme.

\subsection{Numerical implementation}
\label{numerical_implementation}

In the previous section, we discretized the measure by considering an empirical approximation. For a numerical implementation to be tractable, it remains to discretize the time evolution.
Notice already that the IPS method induces no extra computation of the forces, and is therefore unexpensive to implement. However, although the IPS can be parallelized, the processors have to exchange informations at the end of each time step, which can slow down the simulation.

There are several ways to discretize the dynamics~(\ref{e:Pqdyn}) or~(\ref{e:Pdyn}).
Some common schemes used in molecular dynamics are the Euler-Maruyama discretization for~(\ref{e:Pdyn}), and the BBK scheme~\cite{BBK} for~(\ref{e:Pqdyn}).
We refer to~\cite{gaby} for alternative approaches in the field of molecular dynamics.
In the sequel, we will denote by $x^{i,k}$ a numerical approximation
of a realization of $X_{i \Delta t}^k$, with $x=q$ or $x=(q,p)$ depending on the context.

\paragraph*{Euler discretization of the overdamped Langevin dynamics.}
The Euler-Maruyama numerical scheme associated to~(\ref{e:Pdyn})
reads, when taking integration time steps $\Delta t$,
\begin{equation}
\label{space_diff_num}
q^{n+1} = q^{n} - \Delta t \, \nabla V_\lambda (q^{n}) 
+ \sqrt{\frac{2 \Delta t}{\beta}} \, R^{n},
\end{equation}
where $(R^{n})_{n \in \mathbb{N}}$ is a sequence of
independent and identically distributed (i.i.d.) 
$3N$-dimensional standard Gaussian random vectors. 
The numerical convergence of this scheme can be ensured in some circonstances~\cite{gaby}.

\paragraph*{Discretization of the Langevin dynamics.}
When considering an integration time step $\Delta t$, the BBK
discretization of~(\ref{e:Pqdyn}) 
reads componentwise (recall that the underscripts $j$ refer here to the components of a given $x^k \equiv x = (q,p)$), 
\begin{equation}
\label{Langevin_num}
\left \{
\begin{array}{c @{= \ } l}
  p_{j}^{n+1/2} & p_{j}^{n} + \displaystyle{\frac{\Delta t}{2} \left ( - \nabla_{q_{j}}
      V (q^{n}) - \xi \frac{p_{j}^{n}}{m_{j}} 
      + \frac{\sigma_j}{ \sqrt{\Delta t} } R_{j}^{n} \right ) }\\
  q_{j}^{n+1} & \displaystyle{q_{j}^{n} + \Delta t \ \frac{p_{j}^{n+1/2}}{m_{j}}  } \\
  p_{j}^{n+1} & \displaystyle{ \frac{1}{1 + \frac{\xi \Delta t}{2 m_j}}
      \ \left( p_{j}^{n+1/2} - \frac{\Delta t}{2} \nabla_{q_{j}} V
      (q^{n+1}) + \sigma_j \frac{\sqrt{\Delta t}}{2} R_{j}^{n+1} \right)
      }\\ 
\end{array}
\right.
\end{equation}
where the random forcing terms $R_{j}^{n}$ ($j \in \{ 1,\dots,N \} $ is the label 
of the particles, $n$ is the iteration index) are standard i.i.d. Gaussian random variables.
The fluctuation/dissipation relation~(\ref{FDR}) must be corrected so
that the kinetic temperature is correct in the simulations~\cite{gaby}. 
To this end, we set
\begin{equation}
  \label{sigma_num}
  \sigma_j^2 = \frac{2 \xi}{\beta} \left (1 + \frac{\xi \Delta t}{2 m_j} \right ).
\end{equation}
Notice that the relation~(\ref{FDR}) is recovered in the limit $\Delta t \to 0$.

\paragraph*{Discretization of the selection operation}

We consider for example the following discretization of the force exerted on the $k$-th replica on the time interval $[i \Delta t, (i+1)
  \Delta t]$: 
\[
\frac{\partial H^{k,\Delta t}_{\lambda_{i+1/2}}}{\partial\lambda} = \frac{1}{2}
\left ( \frac{\partial H_{\lambda(i \Delta t)}}{\partial\lambda}(x^{i,k})
+ \frac{\partial H_{\lambda((i+1) \Delta t)}}{\partial\lambda}(x^{i+1,k}) \right ).
\]
The mean force is then approximated by
\[
\mathcal{F}^{M, \Delta t}_{\lambda_{i+1/2}} =
\frac{1}{M} \sum_{k = 1}^{M}
\frac{\partial H^{k,\Delta t}_{\lambda_{i+1/2}}}{\partial\lambda}.
\]
To get a time dicretization of the IPS, Process~\ref{p1} is mimicked using the following rules:
\begin{itemize}
\item the time integrals are changed into sums;
\item the selection times are defined as the first discrete times \emph{exceeding} the exponential clocks $\tau^{b/d}$.
\end{itemize}
Further details about the numerical implementation can be found in~\cite{preprint}.
Note that one can find more elaborate methods of discretization of the IPS (see~\cite{moithese}), but this one seems to be sufficient in view of the intrinsic errors introduced by the discretisation of the dynamics.

\section{Applications of the IPS method}
\label{possible_applications}

\subsection{Computation of canonical averages} 

The most obvious application of the IPS method is the computation of phase-space integrals, since an unweighted sample of the target Boltzmann distribution $\mu_{1}$ is generated. The sample obtained can of course be improved by some additional sampling process (according to a dynamics leaving the target canonical measure invariant). This will decorrelate the replicas and may increase the quality of the sample. We refer to~\cite{gaby} for further precisions on sampling methods, to~\cite{gab_these} for some numerical tests assessing the quality of the sample generated as a function the transition time~$T$ and the number of replicas~$M$, and to~\cite{preprint} for an application to pentane.

\subsection{Estimation of free energy difference.}
The free energy of a system is defined as
\[
F = -\beta^{-1} \ln Z
\]
where $Z$ is the partition function $Z = \int \exp (-\beta H(q,p)) \ d q \, d p$. It
cannot be computed with a single sample of $\mu_{\lambda}$.
Only free energy differences can be computed easily. Since the free
energy of certain states is known
(This is the case for perfect gases, or for solids at low temperature~\cite{RS02}),
the free energy of any state can in principle be obtained by an integration between a state
whose free energy is known, and the state of interest. Usual methods
to this end are Thermodynamic
integration~\cite{Kirk35}, the free energy perturbation method~\cite{Zwanz54} and the related Umbrella sampling technique~\cite{TV77}, or Jarzynski's non equilibrium dynamics (also called 'fast growth')~\cite{jarz}.
It is still a matter of debate which method is the most efficient. While some results show that fast growth methods can be competitive in some situations~\cite{HJ01}, other studies disagree~\cite{ODG05}. However, a fair comparison is difficult since the dynamics used may differ (in~\cite{ODG05}, Hamiltonian dynamics are used during the switching process), and more efficient fast growth methods techniques (using, e.g. path sampling~\cite{Sun,YZ04}) are still under investigation.

In the work of Jarzynski~\cite{jarz}, $M$ independent
realizations $(X^{1}_{t},...,X^{M}_{t})$ of a bare out of equilibrium dynamic
\eqref{e:fdyn} are used to compute free energy differences through
\eqref{e:jar-rel}, with the estimator
\[
\Delta \hat{F}_{J} = -\frac{1}{\beta} \ln \left (
\frac{1}{M}\sum_{k=1}^{M} e^{-\beta W^{k}_{1}} \right ).
\]
An alternative (yet similar) estimator relying
on the thermodynamical integration \ref{e:it}, which is also considered in~\cite{Sun} (and is implicit in~\cite{jarz}) is
\[
\Delta \hat{F}'_{J} = \int_{0}^{T} \mathcal{F}^{M_{\rm ind}}_{\lambda(t)} \lambda'(t) \, dt,
\]
where
\[
\mathcal{F}^{M_{\rm ind}}_{\lambda(t)} = \mu^{M_{\rm ind}}_{\lambda(t)}
\left ( \frac{\partial
H_{\lambda(t)}}{\partial\lambda} \right ), \quad \textrm{with} \quad
\mu^{M_{\rm ind}}_{\lambda(t)}(dx) = \frac{\sum_{k=1}^{M} \delta_{X^{k}_{t}}(dx) \,
{\rm e}^{-\beta W^{i}_{t}}}{\sum_{k=1}^{M} {\rm e}^{-\beta W^{i}_{t}}}.
\]
However, both estimators $\Delta \hat{F}_{J}$ and $\Delta
\hat{F}'_{J}$ suffer from the fact that only
a few values of $W_t^i$ are really important. Indeed, because of the
exponential weighting, only the lower tail
of the work distribution is taken into account. The quality of the
estimation then relies on those rare values,
which may be a problem in practice~(see e.g.~\cite{ODG05}).

In the case of interacting replicas, we use similarly
\[
\Delta \hat{F}_{\rm IPS} = \int_{0}^{T} \mathcal{F}^{M}_{\lambda(t)} \lambda'(t) \, dt,
\]
which shares by Proposition~\ref{p:ips} the same statistical properties as $\Delta
\hat{F}_{J}$: $\Delta
\hat{F}_{\rm IPS}$ is asymptotically normal with bias and variance of
order $M^{-1}$, and  the estimator ${\rm e}^{-\beta \Delta \hat{F}_{\rm IPS}}$ is
unbiased estimator of ${\rm e}^{-\beta \Delta F}$.

% %
% %   !!!!! Est ce que l'espérance est bien placée?? Cf. je voyais plutôt 
% %              e^{-\beta \mathbb{E}(W_T)} = e^{-\beta \Delta F}
% %
% %------------------------------------------------------
% %                A REVOIR ENSEMBLE !!
% %------------------------------------------------------
% This means that the Jarzynski's equality still holds when considering the work computed using an IPS dynamics instead of the usual dynamics:
% \[
% \mathbb{E}(e^{- \beta W^{\rm IPS}_{T} } ) = \mathbb{E}(e^{- \beta \int_{0}^{T} \mathcal{F}^{M}_{t} \lambda'(t) \, dt } ) = {\rm e}^{-\beta \Delta F}.
% \]

Our numerical comparisons often turned out to give similar free energy estimations for the IPS method and the standard Jarzynski method. However, we have mostly considered the issue of pure energetic barriers, where the difficulty of sampling comes from overcoming a {\it single high} barrier. The observed numerical equivalence may be explained by the fact that the selection mechanism in the IPS method does not really help to {\it explore} those regions of high potential energy.

However, when the sampling difficulties also come from barriers of more \emph{entropic} nature ({\it e.g.} a succession of very many transition states separated by low energy barriers), the IPS may improve the estimation. Indeed, the selection mechanism helps keeping a statistical amount of replica in the areas of high probability with respect to the local Boltzmann distribution $\mu_{\lambda}$ throughout the switching process (see the numerical example in section~\ref{ss:toy1D}). This relaxation property may be crucial to ensure at each time a meaningful exploration ability. These points are still under investigation.

\subsection{Initial guesses for path sampling}

The problem of free energy estimation is deeply linked with the problem of sampling meaningful transition path. In the IPS method, one can associate to each replica $ X^{k}_{t}$ a \emph{genealogical continuous} path $(X^{k,{\rm gen}}_{s})_{s \in [0,t]}$. The latter is constructed recursively as follows for a replica $k$ (for $0 \leq t \leq T$):
\begin{itemize}
\item at each time $t$, set $X^{k,{\rm gen}}_{t} = X^{k}_{t}$;
\item at each random time $T_n$ when the replica jumps and adopts a new configuration (say of replica $l$), set $(X^{k,{\rm gen}}_{s})_{[0,T_n]}=(X^{l,{\rm gen}}_{s})_{[0,T_n]}$.
\end{itemize}
This path represents the ancestor line of the replica, and is composed of the past paths selected for their low work values. The study of the set of genealogical paths lies outside the scope of this article (see \cite{delmobook} for a discussion in the discrete time case). However, let us mention that for a given $t \in [0,T]$, the set of genealogical paths is sampled, in the limit $M \to \infty$, according to the law of the non-equilibrium paths $(X^{\lambda(s)}_s)_{s\in[0,t]}$ weighted by the factor $e^{-\beta W_{t}}$ (with statistical properties analogous to those of proposition~\ref{p:ips}). These paths are thus typical among non-equilibrium dynamics of those with non-degenerate work. Therefore, they might be fruitfully used as non-trivial initial conditions for more specialized path sampling techniques (as e.g.~\cite{YZ04}).

%%%%%%%%%%%%%%%%%%%%%%%%%%%%%%%%%%%%%%%%%%%%%%%%
%
%   Applications
%
%%%%%%%%%%%%%%%%%%%%%%%%%%%%%%%%%%%%%%%%%%%%%%%%

\section{Numerical simulations}
\label{applications}

\subsection{A toy example}
\label{ss:toy1D}

\begin{figure}
\centering
\includegraphics[height=5cm,width=5.9cm]{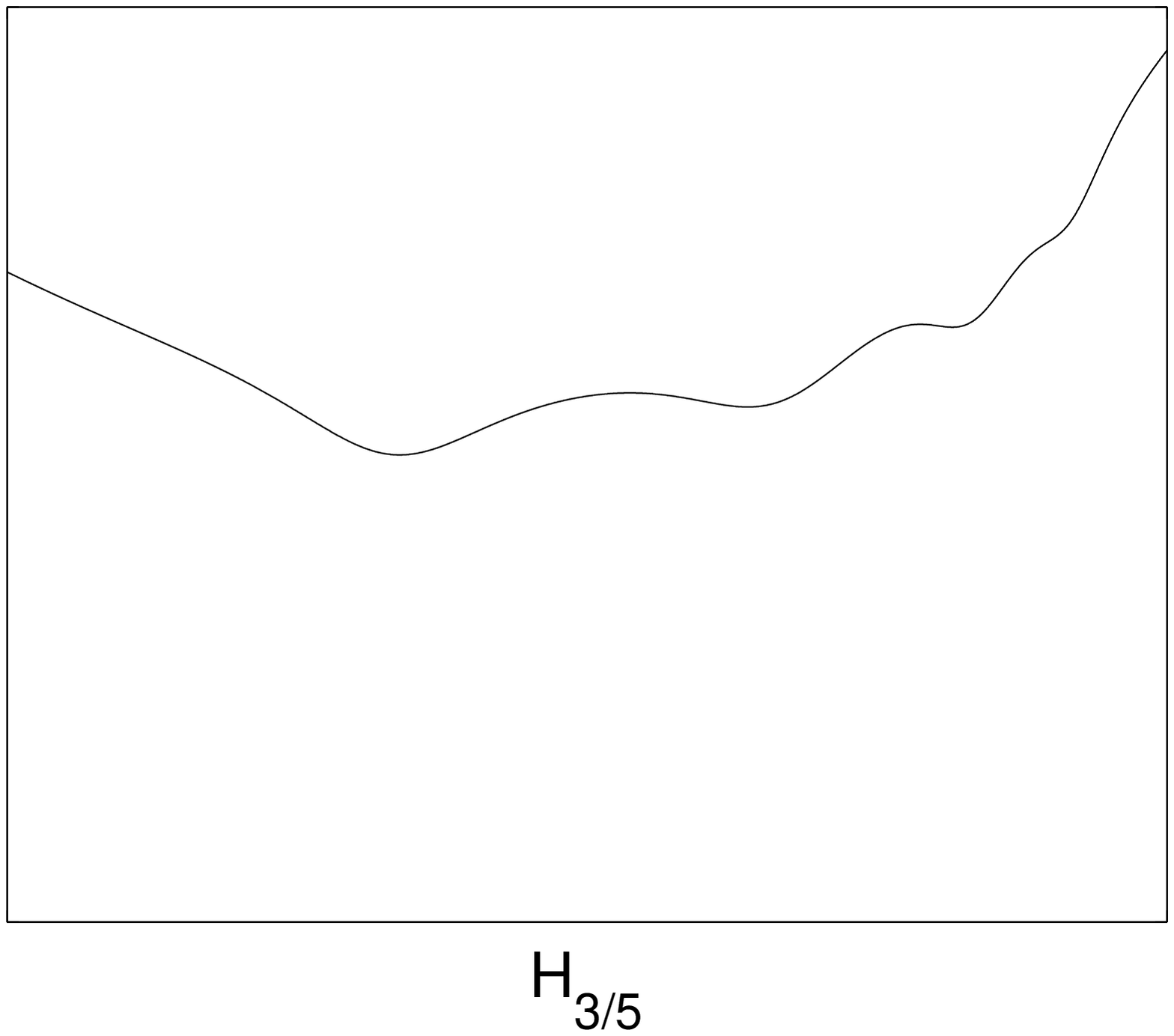}
\includegraphics[height=5cm,width=5.9cm]{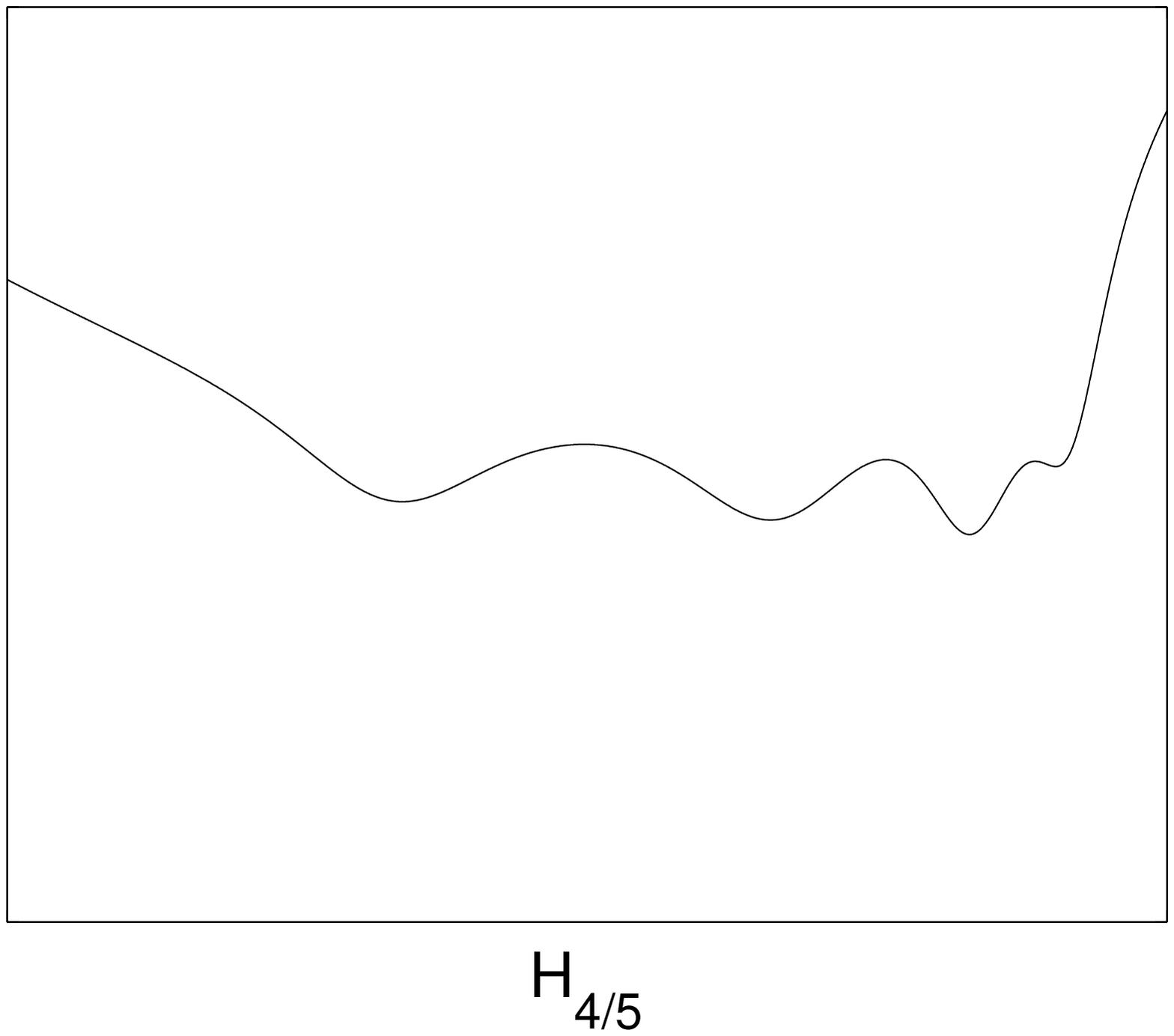}
\includegraphics[height=5cm,width=5.9cm]{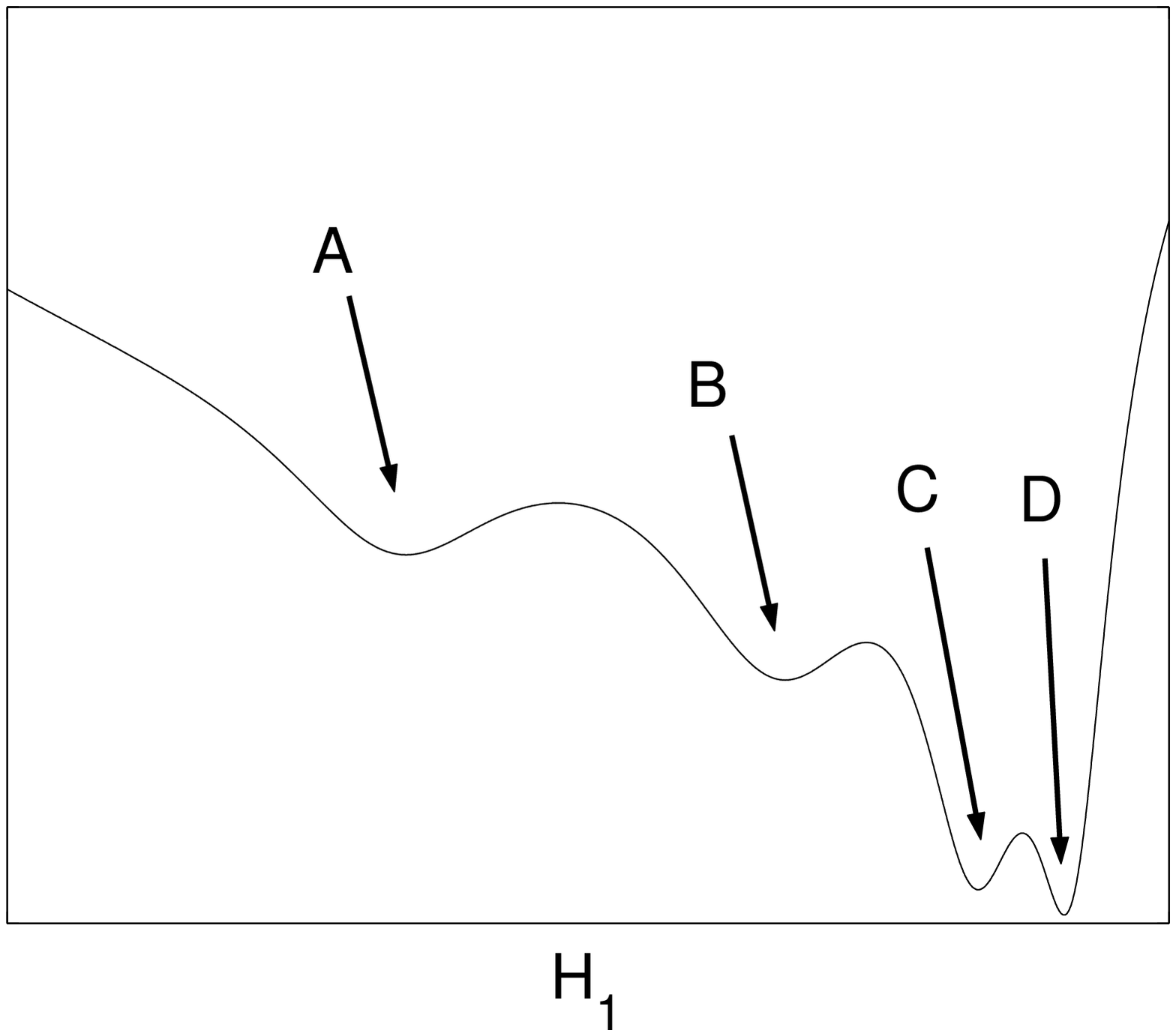}
\caption{Plot of some Hamiltonian functions, as defined by~(\ref{1Dtoy}).}
\label{f:hamilton}
\end{figure}

Consider the following family of Hamiltonians $(H_{\lambda})_{\lambda \in [0,1]}$:
\begin{equation}
\label{1Dtoy}
H_{\lambda}(x) = \frac{x^{2}}{2} + \lambda Q_{1}(x) +
\frac{\lambda^{2}}{2} Q_{2}(x)
+ \frac{\lambda^{3}}{6} Q_{3}(x)+\frac{\lambda^{4}}{24} Q_{4}(x)
\end{equation}
with
\[
Q_1(x)=\frac{-1}{8 x^{2}+1 }, \quad Q_2(x)=\frac{-4}{8 (x-1)^{2}+1 },\\
\]
\[
Q_3(x)=\frac{-18}{32(x-3/2)^{2}+1 }, \quad
Q_4(x)=\frac{-84}{64(x-7/4)^{2}+1 }.\\
\]
Some of those functions are plotted in Figure~\ref{f:hamilton}.
This toy one-dimensional model is reminiscent of the
typical difficulties encountered when $\mu_{0}$ is very different from $\mu_{1}$.
Notice indeed that several transitional metastable states (denoted $A$ and $B$ in Figure~\ref{f:hamilton}) occur in the canonical distribution when going from $\lambda = 0$ to $\lambda = 1$. The probability of presence in the basins of
attraction of the main stable states of $H_{1}$ ($C$ and $D$ in Figure~\ref{f:hamilton}) is only effective when $\lambda$ is close to $1$.

\begin{figure}[h]
\centering
\includegraphics[height=4.6cm,width=5.9cm]{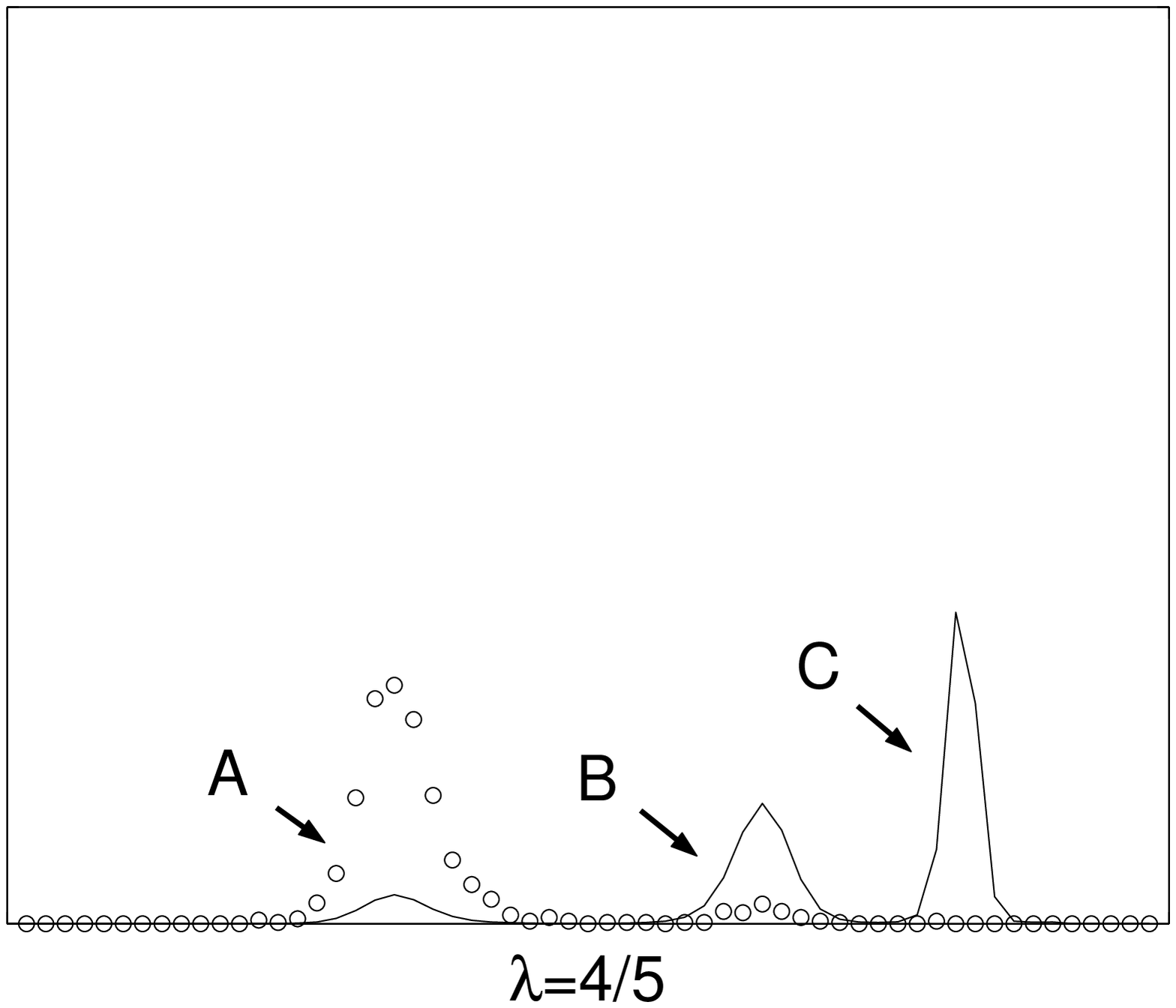}
\includegraphics[height=4.6cm,width=5.9cm]{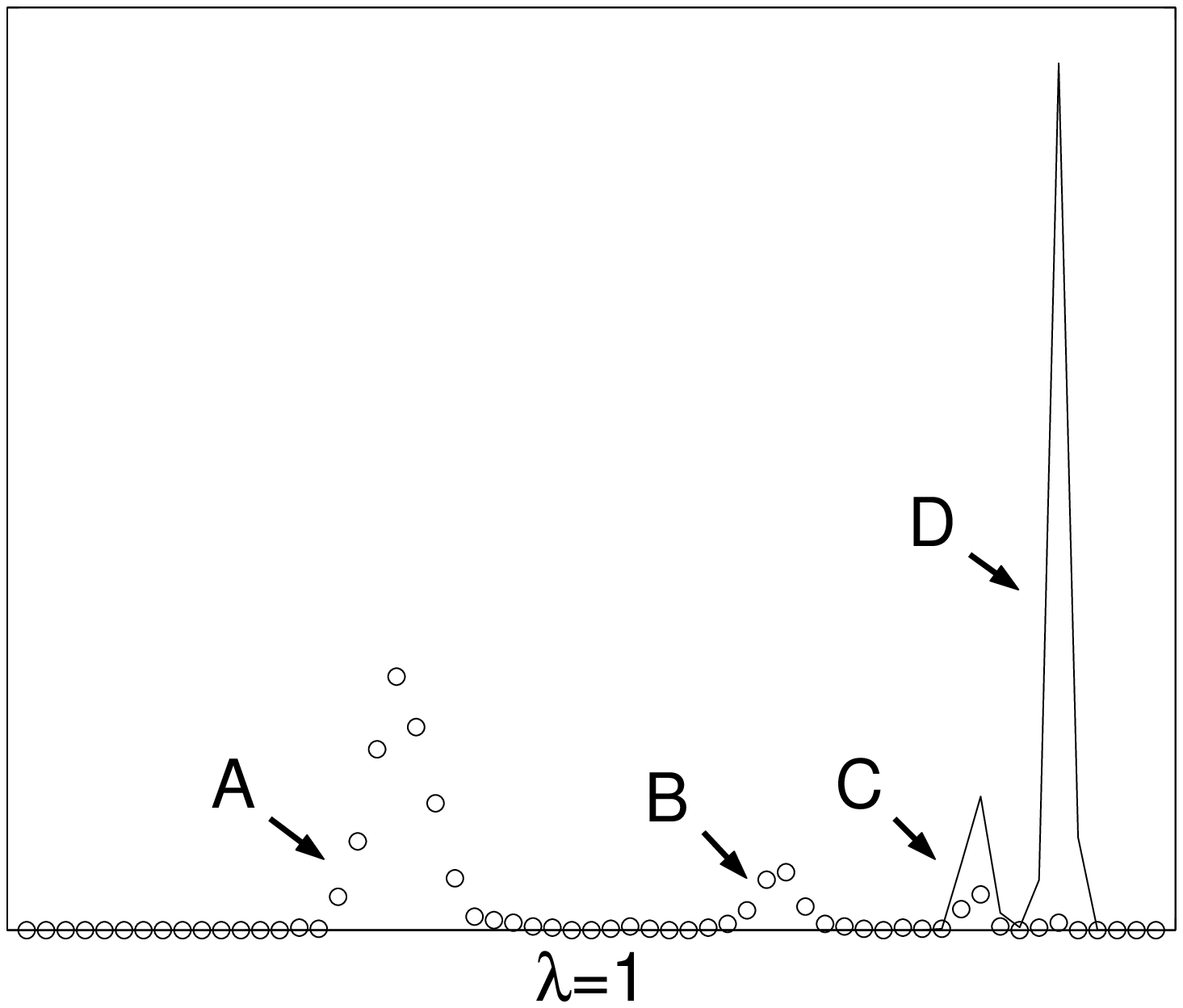}
\caption{Empirical densities (in dots) obtained using independant replicas.}\label{f:indeps}
\end{figure}

\begin{figure}[h]
\centering
\includegraphics[height=4.6cm,width=5.9cm]{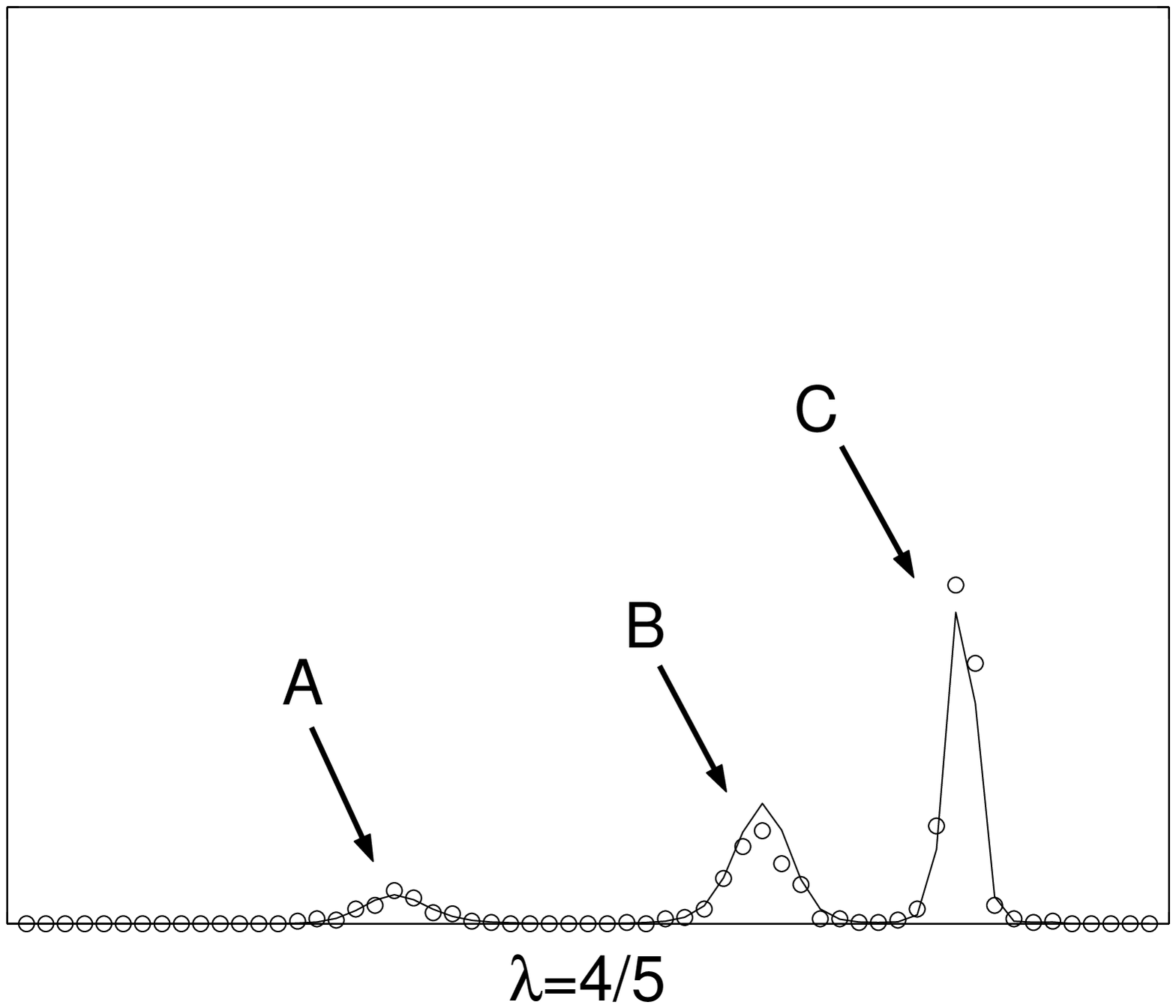}
\includegraphics[height=4.6cm,width=5.9cm]{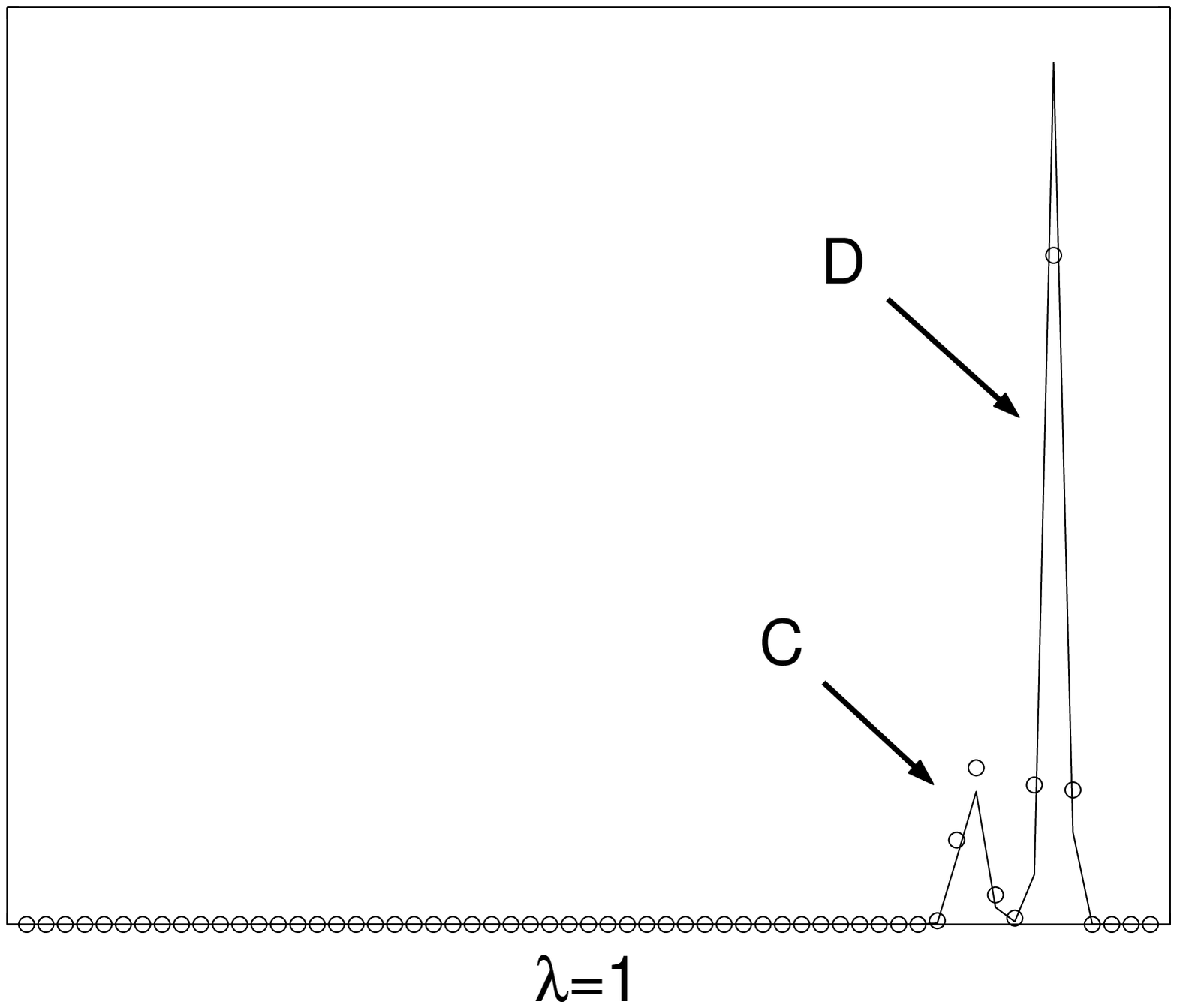}
\caption{Empirical densities (in dots) obtained using interacting replicas.}\label{f:inter}
\end{figure}

Simulations were performed at $\beta = 13$ with the overdamped Langevin dynamics~\eqref{e:Pdyn}, and the above Hamiltonian family~(\ref{1Dtoy}). The number of replicas was $M=1000$, the time step $\Delta t = 0.003$, and $\lambda$ is considered to be linear: $\lambda(t) = t/T$.
Figure~\ref{f:indeps} presents the distribution
of replicas during a slow out of equilibrium plain dynamic:
$T=30$. Figure~\ref{f:inter} presents the distribution
of replicas during a faster dynamics with interaction: $T=15$.

When performing a plain out of equilibrium dynamics (even 'slow') from $\lambda=0$ to $\lambda=1$, almost all replicas are
trapped by the energy barrier of these transitional metastable states (see Figure~\ref{f:indeps}). In the end, a very small (almost null) proportion of replicas have performed interesting paths associated with low values of virtual work $W$. When
using~\eqref{e:FK} to compute thermodynamical quantities, these replicas bear
almost all the weight of the degenerate sample, in view of the exponential weighting.
The quality of the result therefore depends crucially on these rare values.

On the contrary, in the interacting version, the replicas can perform jumps in the configuration space thanks to the selection mechanism, and go from one metastable basin to
another. In our example, as new transition states appear, only few
clever replicas are necessary to attract
the others in good areas (see Figure~\ref{f:inter}). In the end, all
replicas have the same weight, and the sample is not degenerate. Notice also that the final empirical distribution is fairly close to the theoretical one.

We have also made a numerical estimation of the error of the free energy estimation,
with 40 realizations of the above simulation. The results are presented in Table~\ref{t:errorr}, and show
an important reduction of standard deviation and bias up to a factor $2$ when using the IPS method.

\begin{table}
\centering
\begin{tabular}{|c|c|c|}
\hline
Method & Bias & Variance \\
\hline
Plain  & $+0.25$  & $0.19$  \\
\hline
Interacting  &  $+0.15$ & $0.10$ \\
\hline
\end{tabular}
\caption{Error in free energy estimation.}\label{t:errorr}
\end{table}

\subsection{Gradual Widom insertion}

We present here an application to the computation of the chemical potential of a soft sphere fluid. This example was considered in~\cite{HJ01,ODG05} for example. We consider a two-dimensional (2D) fluid of volume $|\Omega|$, simulated with periodic boundary conditions, and formed of $N$ particles interacting via a pairwise potential $V$.
The chemical potential is defined, in the NVT ensemble, as
\begin{equation}
\label{chemical_pot}
\mu = \frac{\partial F}{\partial N},
\end{equation}
where $F$ is the free-energy of the system. Actually, the kinetic part of the partition function $Z$ can be straightforwardly computed, and accounts for the ideal gas contribution $\mu_{\rm id}$. In the large $N$ limit, the chemical potential can be rewritten as~\cite{FS}
\[
\mu = \mu_{\rm id} + \mu_{\rm ex},
\]
whith
\[
\mu_{\rm id} = -\beta^{-1} \ln \left ( \frac{|\Omega|}{(N+1)\Lambda^3} \right ),
\]
where $\Lambda$ is the ``thermal de Broglie wavelength'' $\Lambda = h (2 \pi m
\beta^{-1})^{-1/2}$ (with $h$ Planck's constant). The excess part $\mu_{\rm ex}$ is 
\begin{equation}
\label{muex_ref}
\mu_{\rm ex} = -\beta^{-1} \ln \left ( \frac{\int_{\Omega^{N+1}} \exp( - \beta V(q^{N+1}) ) \, dq^{N+1} }{|\Omega| \int_\Omega \exp( - \beta V(q^N) ) \, dq^N } \right ),
\end{equation}
where $V(q^N)$ is the potential energy of a fluid composed of $N$ particles.
We restrict ourselves to pairwise interactions, with an interaction potential $\Phi$. Then, $V(q^N) = \sum_{1 \leq i < j \leq N} \Phi(|q_i-q_j|)$.
Setting $\pi(q^N) = \exp( - \beta V(q^{N}) )$ and $\Delta V(q^N,q) = V(q^{N+1})-V(q^N)$ with $q^{N+1} = (q^N,q)$, it follows
\begin{equation}
\label{muex}
\mu_{\rm ex} = -\beta^{-1} \ln \left ( \frac{1}{|\Omega|} \int_\Omega {\rm e}^{-\beta \Delta V(q,q^N)} d\pi(q^N) \, dq \right ).
\end{equation} 
The formula~(\ref{muex}) can be used to compute the value of chemical potential using stochastic methods such as the free energy perturbation (FEP) method~\cite{Zwanz54}. In this case, we first generate a sample of configurations of the system according to $\pi$, and then evaluate the integration in the remaining $q$ variable by drawing positions $q$ of the remaining variable uniformly in $\Omega$.

Another possibility is to use fast growth methods, resorting to the following parametrization
\begin{equation}
\label{widom_family}
H_\lambda(q^{N+1},p^{N+1}) = \sum_{i=1}^{N+1} \frac{p_i^2}{2m} + V_\lambda(q^{N+1}) = \sum_{i=1}^{N+1} \frac{p_i^2}{2m} + V(q^N) + \lambda \Delta V(q^N,q).
\end{equation}
In this case, the interactions of the remaining particle are progressively turned on.

\begin{figure}[h]
\includegraphics[width=6.5cm]{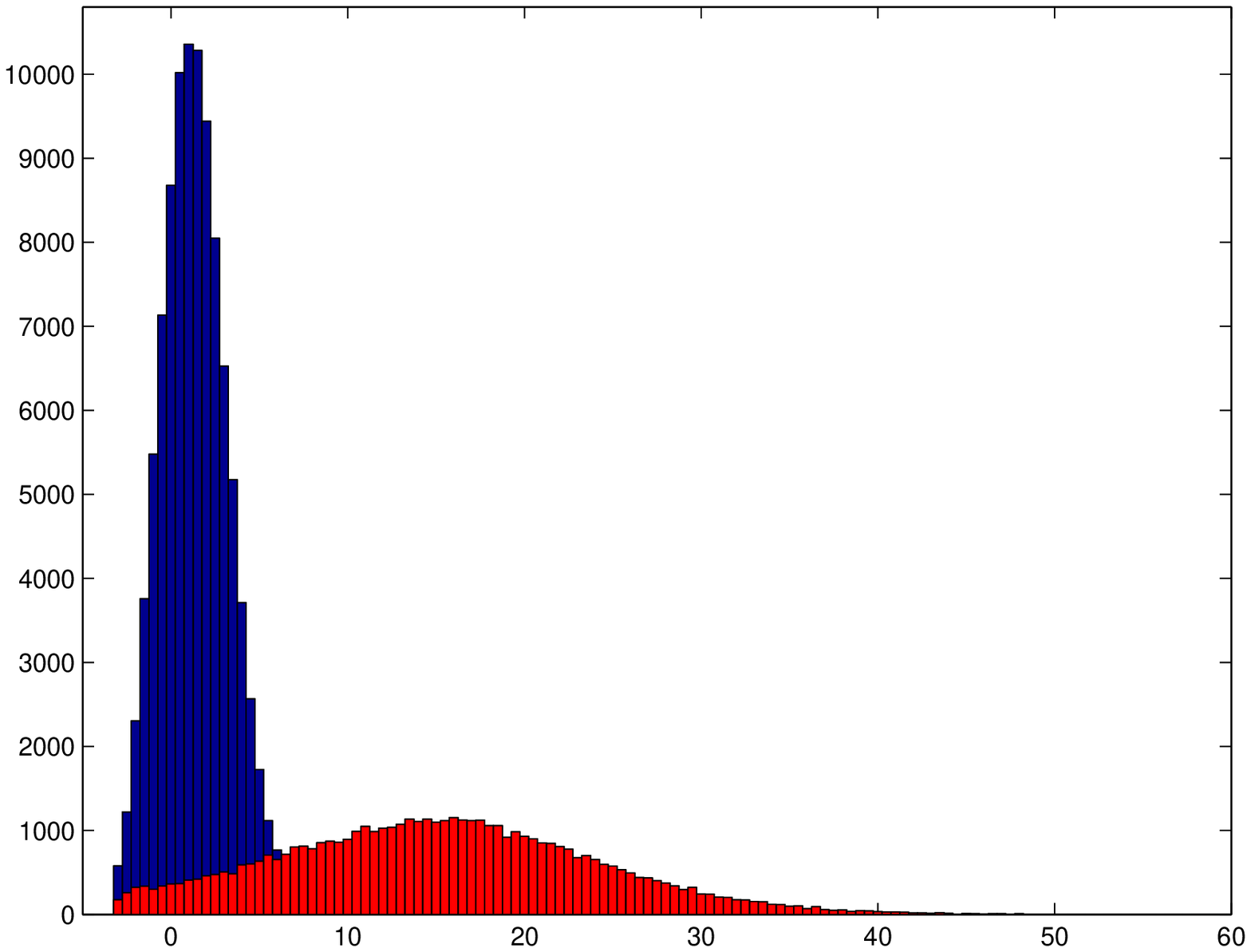}\hfill
\includegraphics[width=6.5cm]{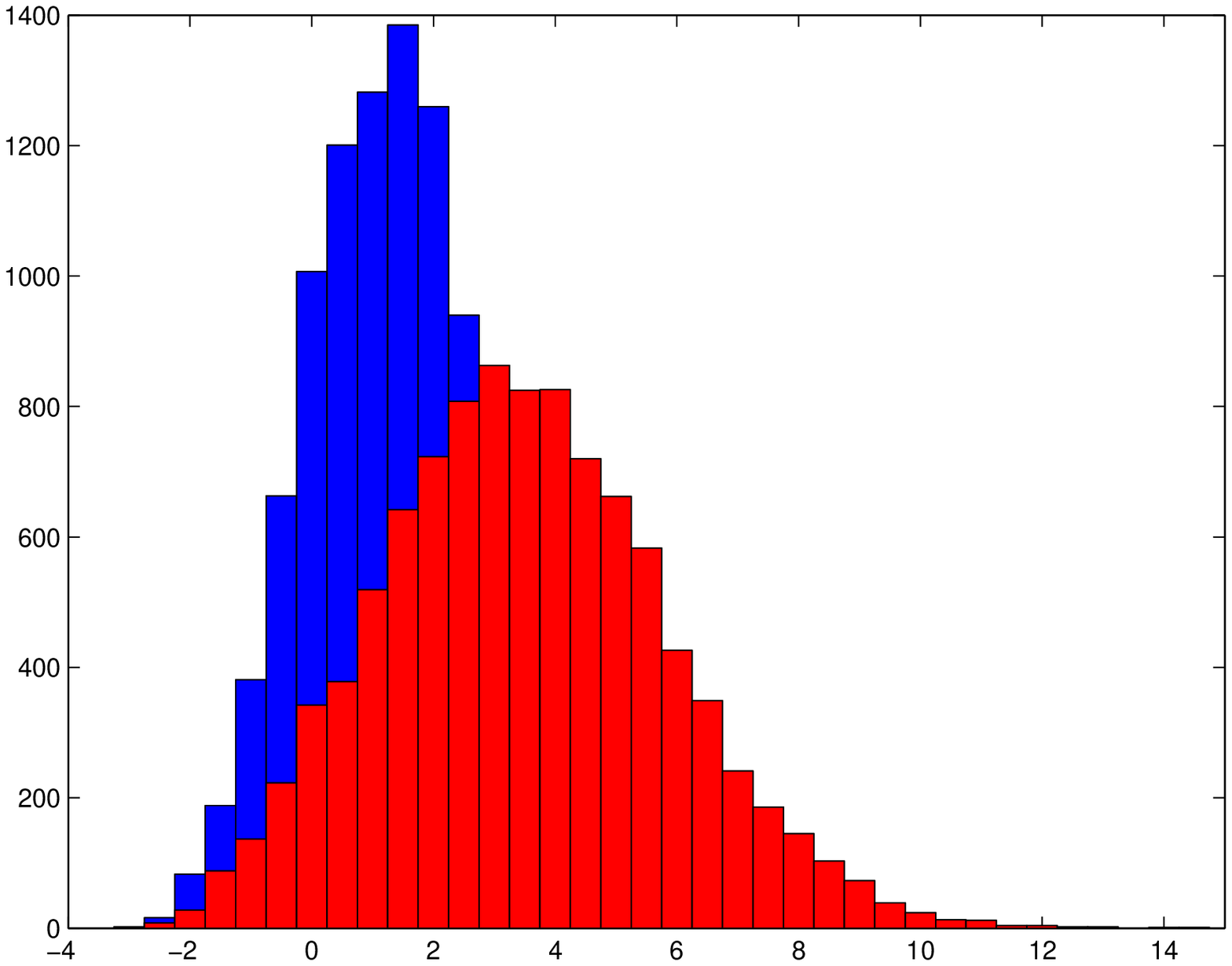}
\caption{ \label{widom_comp} Left: Comparison of the work distribution  for $T=1$. Right: Comparison of the work distributions for $T=10$. The IPS results appear in darker colors. Notice that the bare simulated annealing process work distributions come closer to a gaussian shape, and has a reduced variance as $T$ increases, as expected. }       
\end{figure}

As in~\cite{HJ01,ODG05}, we use a smoothed Lennard-Jones potential in order to avoid the singularity at the origin (Let us however note that, once the particle is inserted, it is still possible to change all the potentials to Lennard-Jones potentials, and compute the correponding free-energy difference). The Lennard Jones potential reads
\[
\Phi_{\rm LJ}(r) = 2 \epsilon \left ( \frac{1}{2} \left (\frac{\sigma}{r} \right )^{12} - \left (\frac{\sigma}{r} \right )^6 \right ),
\]
and the modified potential is
\[
\Phi(r) = \left \{ \begin{array}{ll}
a-br^2, & \quad 0 \leq r \leq 0.8 \, \sigma, \\
\Phi_{\rm LJ}(r) + c(r-r_{\rm c}) - d, & \quad 0.8 \, \sigma \leq r \leq r_{\rm c}, \\
0, & \quad r \geq r_{\rm c}.
\end{array} \right.
\]
The values $a,b,c$ are chosen so that the potential is $C^1$. The distance $r_{\rm c}$ is a prescribed cut-off radius. We considered the insertion of a particle in a 2D fluid of 25 particles, at a density $\rho \sigma^3= 0.8$, with $r_{\rm c} = 2.5 \, \sigma$, $\beta \epsilon = 1$, $\Delta t = 0.0005$, and a schedule $\lambda(t) = t/T$ where $T$ is the transition time. The results are presented in Table~\ref{res_widom}, for different transitions times, but at a fixed computational cost, since $M T$ is constant. Some work distributions are also depicted in Figure~\ref{widom_comp}. A reference value was computed using FEP, with $10^8$ insertions, done by running $M=10^3$ independent Langevins dynamics for the system composed of $N$ particles, for a time $t_{\rm FEP}=50$ (after an initial thermalization time to decorrelate the systems), and inserting one particle at random after each time-step. The reference value obtained is $\mu_{\rm ex} = 1.32 \ k_{\rm B}T$ ($\pm 0.01 \ k_{\rm B}T$).

\begin{table}
\centering
\begin{tabular}{|c|c c c c|}
\hline
Method & $M=10^5$  & $M=5 \times 10^4$ & $M=2 \times 10^4$ & $M=10^4$ \\
 & $T=1$ & $T=2$ & $T=5$ & $T=10$ \\
\hline
SA & 1.32 & 1.29 & 1.33 & 1.36 \\
\hline
IPS & 1.37 & 1.35 & 1.29 & 1.34 \\
\hline
\end{tabular}
\caption{\label{res_widom} Free energy estimation for one realisation of each method, depending on the switching time $T$ and the number of replicas $M$ used, keeping $M T$ constant. The reference value obtained through FEP is $\mu_{\rm ex} = 1.31 \ k_{\rm B}T$ ($\pm 0.01 \ k_{\rm B}T$). Notice that the results are quite comparable.}
\end{table}

%\begin{figure}[h]
%\includegraphics[width=6.5cm]{images/widom_IPS.eps}\hfill
%\includegraphics[width=6.5cm]{images/widom_SA.eps}
%\caption{ \label{widom_figures} Repartition of the work values for $M=10^5$ and $T=1$. Left: IPS work distribution. Right: bare simulated annealing process. Notice that the IPS work distribution has indeed a gaussian shape. Note also the different scale of the axes.}
%\end{figure}

As can be seen from the results in Table~\ref{res_widom}, the IPS algorithm has a comparable accuracy to Jarzynski's estimates. However, the work distribution is very different, and has a gaussian shape for all switching rates considered, whereas the work distribution obtained through the fast growth method are much wider (see in particular Figure~\ref{widom_comp} (Left)), so that the relevant part of the work distribution (the lower tail) is only of small relative importance. Of course, the width of this distribution decreases as the transition is made slower.

%%%%%%%%%%%%%%%%%%%%%%%%%%%%%%%%%
%
%  Conclusion et perspectives
%
%%%%%%%%%%%%%%%%%%%%%%%%%%%%%%%%%

%%%%%%%%%%%%%%%%%%%%%%%%%%%%%
%
% Appendice
%
%%%%%%%%%%%%%%%%%%%%%%%%%%

\section*{Appendix : Pure jump processes}

Consider a Markov process $X_{t}$ of infinitesimal generator
\[
J_{t}(f)(x) = \int (f(y)-f(x)) \alpha_{t}(x)d\mu_{t}(y),
\]
where for each time $t$, $\alpha_{t}$  is a bounded positive function and $\mu_{t}$ a probability measure. Denote by $(T_{n})_{n \geq 1}$ the jump times (with $T_{0}=0$), and $(\tau_{n})_{n \geq 1}$ independant clocks of exponential law of mean $1$.
Then the system evolves according to the following stochastic rules:
\begin{itemize}
\item  the jump times are defined by
$$\int_{T_{n}}^{T_{n+1}}\alpha_{t}(X_{t})dt=\tau_{n+1};$$
\item at jump times, the process jumps to a configuration chosen according to the probability measure
$d\mu_{T_{n+1}}(y)$.
\end{itemize}
\end{document}